%% file: main.tex
\documentclass[10pt,conference,letterpaper]{IEEEtran}
\usepackage{cite}
\usepackage{amsmath,amssymb,amsfonts}
\usepackage{dsfont}
\usepackage{graphicx}
\usepackage{textcomp}
\usepackage[table]{xcolor}
\usepackage{booktabs}
\usepackage{colortbl}
\usepackage{multirow}
\usepackage{float}
\usepackage[hidelinks]{hyperref}
\IEEEoverridecommandlockouts
\makeatletter
\def\@IEEEauthorblockAtopspace{0.5ex}
\makeatother
\def\BibTeX{{\rm B\kern-.05em{\sc i\kern-.025em b}\kern-.08em
    T\kern-.1667em\lower.7ex\hbox{E}\kern-.125emX}}
\begin{document}
\bstctlcite{IEEEtran:BSTcontrol}

%\title{Optimizing Satellite Downlinking through Clear-Weighted Bit Allocation}
\title{
Clear-Weighted Bit Allocation for Satellite Downlinks
}

\author{
\IEEEauthorblockN{Alireza Furutanpey\textsuperscript{1,2,*}\thanks{\textsuperscript{*}Corresponding author: \href{mailto:a.furutanpey@dsg.tuwien.ac.at}{a.furutanpey@dsg.tuwien.ac.at}},
Qiyang Zhang\textsuperscript{3},
Yujie Huang\textsuperscript{4},
Philipp Raith\textsuperscript{1,2},
Schahram Dustdar\textsuperscript{2,5}}
\IEEEauthorblockA{\textsuperscript{1}Coovally, Barcelona, Spain \quad
\textsuperscript{2}Technical University of Vienna, Austria \quad
\textsuperscript{3}Peking University, China \\
\textsuperscript{4}Beijing University of Posts and Telecommunications, China \quad
\textsuperscript{5}ICREA, Barcelona, Spain}
}

\maketitle

\begin{abstract}
\input{abstract}
\end{abstract}


\input{sections/introduction}
\input{sections/background}
\input{sections/method}
\input{sections/scheduling}
\input{sections/evaluation}
\input{sections/related_work}
\input{sections/conclusion}

\clearpage
\bibliographystyle{IEEEtran}
\phantomsection\label{refstart}% references start page, read by check_formatting.sh
\IEEEtriggeratref{19}
\bibliography{references}

\end{document}

%% file: abstract.tex
Earth-observation satellites capture more imagery than intermittent ground contacts can transmit. Onboard systems threshold a cloud detector, discard frames or tiles, and compress the survivors with a fixed codec. On expert-labeled imagery, these rules remove more than one-fifth of clear pixels, primarily through detector false positives. We train a neural codec with a clear-probability-weighted reconstruction loss, reallocating coded bytes from clouds to clear ground without requiring or transmitting a cloud map onboard. Each capture is encoded into a resumable base layer and a dependent refinement layer, while clear content is estimated from features produced by the encoder. At each contact, we causally rank arrived layers using estimated clear content, unfinished bytes, deadline slack, and aggregate deadline pressure. The scheduler serves base and computational deadlines, bounds stored residual bytes, and resumes interrupted packets. We evaluate the onboard-to-downlink pipeline using real entropy-coded bytes, orbit-derived interruptible contact capacities, and measured service time and energy on resource-constrained embedded accelerators. Clear-weighted codecs require up to 47.8\% fewer bytes than learned-compression baselines at matched clear-region quality. The optimized encoder consumes less time and energy than one pass of the cloud detector used by the frame-discard rules. Relative to fixed two-stage service on the same streams, our scheduler more than doubles deadline-full clear-content delivery for the interrupted combined cohort, reaches 83.6\% of a certified clairvoyant upper bound, and exceeds replayed reference orders in deadline-usable delivery.

%% file: sections/introduction.tex
\section{Introduction}\label{sec:intro}
Earth-observation satellites acquire imagery throughout an orbit but transmit only during intermittent ground contacts. With contact number and duration fixed, higher-resolution sensors produce more pixels per capture, and larger constellations produce more captures~\cite{oec,DBLP:conf/infocom/ZhangYXZZMXDW24,earthplus}. An onboard system must allocate source bits within each capture and downlink bytes among captures before their deadlines while respecting processing, storage, and downlink limits~\cite{kodan,fool,deepspace}.

Prior onboard Earth-observation systems discard frames, select tiles, fill image regions, order payloads, or control receiver-side reconstruction from image content, cloud coverage, or task value~\cite{oec,DBLP:conf/infocom/ZhangYXZZMXDW24,kodan,earthplus,deepspace,scocloud}. Their authors compute these estimates outside the source codec and do not use them to allocate source bits between clear and cloudy regions. 
Studies of prior learned compression for satellite imagery optimize latent representations and entropy models while holding the acquired images and delivery schedule fixed, and therefore do not compare downlink byte allocations across captures~\cite{cosmic,phisat2,terracodec,rs_llic,molliere2025}.
Conventional onboard systems estimate cloud coverage, apply a threshold, and discard entire frames or tiles before source coding~\cite{cloudscout,earthplus,kodan}. When an onboard system discards a frame or tile, it removes all clear-ground pixels from that frame or tile from encoding and the downlink queue. When the classification is wrong, the system removes clear ground that a correct decision would retain. With atomic payloads, no part of a capture can be reconstructed until every byte arrives.
Even when payloads are layered, fixed two-stage service reserves contact capacity for base layers whose deadlines have arrived and admits refinement bytes only afterward. It does not jointly account for unfinished work at both deadlines.

During training, we use pixelwise cloud probabilities from a frozen detector to assign larger reconstruction-error weights to likely clear ground. We confine that detector to training. During operation, we estimate each capture's clear-pixel count using a convolutional readout of the already computed hypersynthesis features. We use this estimate as the codec-native semantic value for scheduling, replacing the separate operational detector, and produce a resumable base layer and a dependent refinement layer.
At each contact, we compute a deadline-pressure max weight (DPMW) for every arrived layer from its capture's codec-native value, unfinished item bytes, separate base and full deadline terms, and the current queue state. We serve arrived layer bytes in decreasing weight order. We assign earlier deadlines to base layers than to complete items, limit storage, and retain each partial packet's byte offset so service can resume at a later contact.

We isolate source allocation by holding the service rule and workload fixed while comparing clear-weighted and cloud-agnostic encodings, and isolate service orders by comparing schedulers over the same clear-weighted byte streams and workload. At validation operating points matched for clear-region reconstruction quality, we measure 21\% fewer transmitted bytes, including layer headers, with clear-weighted progressive encoding than with cloud-agnostic encoding from the same codec architecture. Clear-weighted encoding raises deadline-full delivery across orbit-derived contact capacities when both streams use fixed two-stage service. For the interrupted combined cohort across the trained codecs in the held-out contact workload, DPMW more than doubles deadline-full delivery relative to fixed two-stage service. It reaches 83.6\% of a certified clairvoyant upper bound and exceeds the four replayed reference orders in deadline-usable delivery.
For the primary Orin Nano Super configuration, we measure the source-coding service time from data preparation through entropy coding and find it below the computational deadline derived from reported capture intervals~\cite{kodan}. We provide the implementation and evaluation scripts in the accompanying repository\footnote{\url{https://github.com/rezafuru/Clear-Weighted-Bit-Allocation}}.

Our main contributions are:
\begin{itemize}
  \item We develop clear-weighted learned source coding by using pixelwise cloud probabilities during training to allocate source bits toward clear ground.
  \item We design a progressive codec-network interface that represents each capture with a resumable base layer, a dependent refinement layer, and a codec-native semantic value estimating clear-pixel count.
  \item We develop causal two-deadline scheduling that ranks arrived layers by codec-native value, unfinished work, and deadline slack.
\end{itemize}
%While we do not consider the coding-path optimizations an independent research contribution, we release the optimized entropy-coding integration in a separate repository\footnote{\url{https://anonymous.4open.science/r/RANSKit-CD5F}}. e.
%While we do not consider the coding-path optimizations an independent research contribution, we release the optimized entropy-coding integration in a separate repository\footnote{\url{https://anonymous.4open.science/r/RANSKit-CD5F}}. To the best of our knowledge, no prior satellite-imagery system combines training-time clear-weighted source allocation with causal multi-contact service of resumable base and refinement bytes using a codec-native semantic value.
While we do not consider the engineering work to optimize entropy coding and execution paths a research contribution, we release these in a separate repository\footnote{\url{https://github.com/rezafuru/RANSKit}} to reduce implementation confounds that otherwise favor systems-centric methods.
To the best of our knowledge, no prior satellite-imagery system considers training-time clear-weighted source allocation with causal multi-contact service of resumable base and refinement bytes using a codec-native semantic value.

%% file: sections/background.tex
\section{Background and System Model}\label{sec:model}
Captured images arrive between intermittent ground contacts and are persisted in finite onboard storage. At each contact, only the current byte budget is available, and we allocate it causally among layers of arrived objects while unsent bytes remain queued. A usable base reconstruction has an earlier deadline than the full reconstruction, which requires both layers.

We first assign bytes within each image through source coding, then choose which layer bytes to transmit so that bases and complete objects meet their deadlines. Each coded object contains a resumable base layer, a dependent refinement layer, and a codec-native semantic value that is fixed upon arrival.

We measure frozen detector probabilities to weight reconstruction error during codec training and binary training-split cloud masks to supervise the native readout. Offline, we use validation reference masks for operating-point and resource selection and for validation-only comparisons, and test reference masks to construct held-out workloads and score test results. For detector-based baselines, we run the standalone detector.
Let $H$ and $W$ be image height and width, $\mathbf{x}_k\in[0,1]^{H\times W\times3}$ the RGB frame $k$, and $\hat{\mathbf{x}}_k$ its reconstruction. Detector output $q_{ki}\in[0,1]$ is the estimated cloud probability at pixel $i$. The evaluator-only indicators $C_{ki}\in\{0,1\}$ and $G_{ki}\in\{0,1\}$ mark reference-clear and reference-cloud pixels. For AllClear, we construct the reference regions from cloud-mask band 2~\cite{allclear}. We set $G_{ki}=1$ where the band equals one, $C_{ki}=1$ where it equals zero, and $C_{ki}=G_{ki}=0$ elsewhere. We do not read the separate shadow bands. The unscored pixels remain in the $HW$ denominator of the frame cloud fractions. For CloudSEN12+, we set $G_{ki}=1$ on the expert thick- and thin-cloud classes and $C_{ki}=1$ elsewhere.
\subsection{Cloud-Aware Source Decisions}
The conventional detect-then-drop pipeline thresholds clouds per frame and encodes only the survivors. For this pipeline, we compute the predicted and reference frame cloud fractions as
\begin{equation}
\hat\rho_k=\frac{1}{HW}\sum_i\mathds{1}\{q_{ki}>0.5\},
\qquad
\rho_k=\frac{1}{HW}\sum_iG_{ki},
\label{eq:frame-cloud}
\end{equation}
where $\mathds{1}\{\cdot\}$ is the indicator function. We discard frame $k$ when $\hat\rho_k>\tau$ for a cloud-fraction threshold $\tau$ and compress the remaining frames. We instantiate $\tau$ at the published CloudScout 70\% and Earth+ 50\% frame thresholds, which we respectively call the CloudScout Rule and the Earth+ Rule~\cite{cloudscout,earthplus}. The pixel-removal adaptation of Wang et al. instead removes detector-predicted cloud pixels inside the codec and transmits the pixel mask as a side channel~\cite{wang2024}.
For clear-weighted encoding, we train on every frame with reconstruction error weighted by $1-q_{ki}$.

\begin{figure}[t]
\centering
\includegraphics[width=\columnwidth,trim=1.38pt 4.26pt 3.36pt 4.26pt,clip]{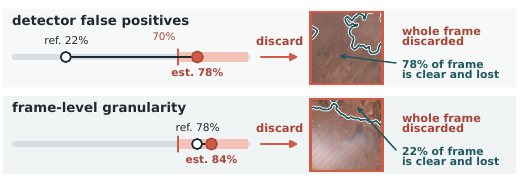}
\caption{Clear-content deletion caused by the CloudScout Rule.}
\label{fig:deletion-mechanisms}
\vspace{-0.75\baselineskip}
\end{figure}

Let $n_k^C=\sum_i C_{ki}$ be the number of scored clear pixels in frame $k$.
Capture-time deletion is
\begin{equation}
\delta(\tau)=
\frac{\sum_k n_k^C\mathds{1}\{\hat\rho_k>\tau\}}{\sum_k n_k^C}
=\delta_{\mathrm{fp}}(\tau)+\delta_{\mathrm{sl}}(\tau),
\label{eq:deletion}
\end{equation}
where
\begin{align}
\delta_{\mathrm{fp}}(\tau)
&=\frac{\sum_k n_k^C\mathds{1}\{\hat\rho_k>\tau,\rho_k\leq\tau\}}
{\sum_k n_k^C}, \nonumber\\
\delta_{\mathrm{sl}}(\tau)
&=\frac{\sum_k n_k^C\mathds{1}\{\hat\rho_k>\tau,\rho_k>\tau\}}
{\sum_k n_k^C}.
\label{eq:deletion-components}
\end{align}
We count clear pixels from detector false positives in $\delta_{\mathrm{fp}}$ and clear pixels inside correctly discarded cloudy frames, which we call clear slivers, in $\delta_{\mathrm{sl}}$ (Fig.~\ref{fig:deletion-mechanisms}). The second term remains positive with perfect frame-level estimates because a frame above the cloud-fraction threshold can still contain clear ground. Clear-weighted encoding has zero capture-time deletion because we encode every frame. Reordering retained layer bytes after coding cannot recover a discarded frame.
\subsection{Performance Measures}
For an emitted payload $b_k$, we compute the coded rate as $r_k=8\lvert b_k\rvert/(HW)$ bits per pixel (bpp). For learned codecs, we count every emitted image-latent and hyperlatent byte string, excluding dimensions, container metadata, and shared decoder state. For JPEG 2000~\cite{jpeg2000}, we count the self-contained codestream after removing its comment marker, including 118 bytes of rate-independent marker segments per frame that the learned codec excludes.
We use likelihood cross-entropy as the training-rate proxy and the emitted bytes for every reported rate-distortion comparison.
We compute the clear-region peak signal-to-noise ratio (PSNR) over the RGB reconstruction error on reference-clear pixels and omit frames without any scored clear pixels from per-frame statistics. Let $\mathcal{K}$ be a deadline-eligible cohort, $\mathcal{A}_B(\mathcal{K})$ its items whose base layer meets its deadline, and $\mathcal{A}_F(\mathcal{K})$ those whose base and refinement layers meet the full deadline. For the base and full deadlines $\theta\in\{B,F\}$, reference-clear delivery is
\begin{equation}
\eta_\theta(\mathcal{K})=
\frac{\sum_{k\in\mathcal{A}_\theta(\mathcal{K})}n_k^C}
{\sum_{k\in\mathcal{K}}n_k^C}.
\label{eq:delivered-clear}
\end{equation}
We call $\eta_B$ deadline-usable delivery and $\eta_F$ deadline-full delivery.
We measure the fused batch-1 service time $S$ from tile preparation through host-to-device transfer, neural encoding, device-to-host transfer, and entropy coding, followed by one accelerator synchronization. We exclude queueing delay, and a batch size of 1 gives zero batch-fill delay. A satellite observes a new multi-tile image every 1--30\,s~\cite{kodan}. Sustained single-processor encoding divides this interval among the tiles, our $256\times256$ frames, leaving milliseconds on wide sensors and seconds on narrow ones. We test the empirical 99th percentile of $S$ against the 33\,ms target inside this span. For a frame-discard rule over $K$ captured frames, we compute the survivor fraction as
\begin{equation}
s(\tau)=\frac{1}{K}\sum_{k=1}^{K}
\mathds{1}\{\hat\rho_k\leq\tau\}.
\label{eq:survivor}
\end{equation}
We derive serial service demand from measured component medians, counting image tiling, host-to-device transfer, and detector time for every captured frame, and adding neural encoding, device-to-host transfer, and entropy encoding for the survivor fraction $s(\tau)$. 
%We measure fused-tail service time separately.

We measure energy from the total module rail over separate active and idle windows. Let $\bar P_{\mathrm{act}}$ and $\bar P_{\mathrm{idle}}$ denote mean active and idle power, $T_{\mathrm{win}}$ the active-window duration, and $n$ the frames completed during that window. We compute the idle-subtracted energy per frame as
\begin{equation}
\mathcal{E}=\max\{0,\bar P_{\mathrm{act}}-\bar P_{\mathrm{idle}}\}
\frac{T_{\mathrm{win}}}{n}.
\label{eq:energy}
\end{equation}

%% file: sections/method.tex
\section{Clear-Weighted Progressive Coding}\label{sec:method}

\begin{figure}[t]
\centering
\includegraphics[width=\columnwidth]{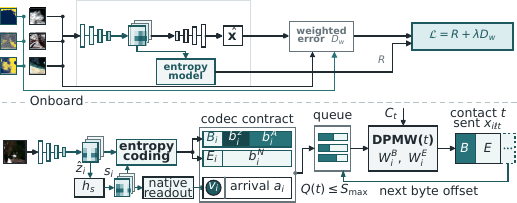}
\caption{Clear-weighted training and causal service of progressive layers.}
\label{fig:clear-weighted-path}
\vspace{-0.75\baselineskip}
\end{figure}

During training, we weigh reconstruction error with a frozen cloud detector. Onboard, we emit resumable base and refinement bytes, along with a codec-native semantic value, for each capture, and serve these layers across contacts in Section~\ref{sec:scheduling} (Fig.~\ref{fig:clear-weighted-path}).

\subsection{Clear-Weighted Training Objective}
\label{sec:method:objective}

We omit frame index $k$ in the per-frame formulation. We estimate cloud probability $q_i$ with a four-level U-Net of base width 64 from Sentinel-2 bands B8, B4, B3, and B2. We replace non-finite reflectance with zero, divide by 10,000, and clip to $[0,1]$. For each clear-weighted codec seed, we freeze the detector checkpoint from that seed's preceding AllClear detector-and-codec training run at distortion weight 300. Detector probabilities and RGB reconstructions use the same $256\times256$ crop without resampling. The clear weight is
\begin{equation}
w_i=1-q_i.
\label{eq:weight}
\end{equation}

We input RGB bands B4, B3, and B2 to the codec. For per-pixel RGB error $e_i=\lVert\mathbf{x}_i-\hat{\mathbf{x}}_i\rVert_2^2/3$, where $\mathbf{x}_i\in[0,1]^3$ is the color triple at pixel $i$, we use normalized clear-weighted distortion. Let $R$ be the likelihood rate proxy, $\lambda>0$ the distortion weight, and $\epsilon=10^{-8}$.
\begin{equation}
D_w=\frac{\sum_iw_ie_i}{\sum_iw_i+\epsilon},
\qquad
\mathcal{L}=R+\lambda D_w.
\label{eq:wmse}
\end{equation}
Let $y$ and $z$ denote the image latent and hyperlatent, with quantized forms $\hat y$ and $\hat z$. Indices $j$ and $m$ enumerate their coordinates. We use $p_{z,m}$ for the modeled hyperlatent probability mass and $p_{y,j}$ for the conditional image-latent probability mass. The rate proxy is
\begin{equation}
\begin{split}
R=-\frac{1}{HW}\bigg[
&\sum_{m}\log_2p_{z,m}(\hat z_m)\\
&+\sum_j\log_2p_{y,j}(\hat y_j\mid\hat z,\mathcal{C}_j)
\bigg],
\end{split}
\label{eq:rateproxy}
\end{equation}
where $\mathcal{C}_j$ is the decoded checkerboard context. We train the entropy-bottleneck quantiles with a separate auxiliary loss. We use detector-derived $w_i$ only during codec optimization and encode each operational frame from RGB alone, transmitting no cloud map.

For a generic binary clear indicator $Z_i$, suppose the detector is calibrated so $w_i=\Pr(Z_i=1\mid V)$ for observation $V$. If reconstruction error is conditionally independent of $Z_i$ given $V$, then
\begin{equation}
\mathbb{E}[w_ie_i]=\mathbb{E}[Z_ie_i].
\label{eq:expected-clear-distortion}
\end{equation}
If calibration and conditional independence hold, we estimate expected squared error on clear pixels from the numerator of $D_w$ and clear area from its denominator. We implement the objective with the continuous score $w_i=1-q_i$, including pixels outside the scored clear and cloud regions.

\subsection{Rate-Allocation Interpretation}

We model latent coordinate $j$ as a separable Gaussian with variance $\nu_j$, distortion $d_j$, and ideal rate $R_j(d_j)=[\tfrac12\log_2(\nu_j/d_j)]^+$, where $[u]^+=\max\{u,0\}$. Let $\tilde w_j$ denote its effective clear weight, averaged over the decoder receptive field and normalized to unit mean. The local allocation is
\begin{equation}
\min_{\substack{\{d_j\}_j\\0<d_j\leq\nu_j\ \forall j}}
\sum_j\left[R_j(d_j)+\lambda\tilde w_jd_j\right].
\label{eq:local-allocation}
\end{equation}
Let $r_j^\star=R_j(d_j^\star)$ denote the minimizing rate. The solution is
\begin{equation}
\begin{split}
d_j^\star&=\min\left\{\nu_j,
\frac{1}{2\ln2\,\lambda\tilde w_j}\right\},\\
r_j^\star&=\left[
\frac{1}{2}\log_2(2\ln2\,\lambda\nu_j\tilde w_j)
\right]^+ .
\end{split}
\label{eq:local-solution}
\end{equation}
For $\tilde w_j=0$, we set $d_j^\star=\nu_j$ and $r_j^\star=0$. When both solutions have positive rate, changing the same coordinate from unit weight to $\tilde w_j$ changes its ideal rate by $\tfrac12\log_2\tilde w_j$. Receptive fields and shared parameters couple coordinates after retraining, so we use the derivation only to predict that larger effective clear weight lowers distortion and raises rate where the positive-rate solution applies.

\subsection{Progressive Codec Interface}

We implement the analysis and synthesis transforms with Cheng-style residual blocks, strided convolutions, and subpixel upsampling~\cite{cheng2020}. The analysis transform $g_a$ maps $\mathbf{x}$ to $y$, and the hyper-analysis transform $h_a$ maps $y$ to $z$. After quantizing $z$, the hyper-synthesis transform $h_s(\hat z)$ produces side parameters for the conditional distribution of $\hat y$. The synthesis transform $g_s$ reconstructs the image from $\hat y$.

We code image latents with a two-pass checkerboard context~\cite{checkerboard2021}. Anchors use only the hyperprior side parameters. Non-anchors additionally use a masked convolution over decoded anchors and separate pointwise networks for Gaussian means and scales, adapting the improved checkerboard context of Fu et al.~\cite{fu2024checkerboard}. The encoder uses a $5\times5$ context over 12 opposite-parity anchors.

We emit the hyperlatent, anchor, and non-anchor streams in dependency order. The base layer contains the hyperlatent and anchors. To decode it, we reconstruct the anchors, fill missing non-anchors with their conditional means, and apply $g_s$. We decode the refinement layer's non-anchor stream to recover the unchanged full reconstruction. We encode each image once, with no codec retraining and no second low-rate payload.

In the network formulation, $i$ indexes coded items rather than pixels. Let $b_i^z$, $b_i^A$, and $b_i^N$ denote the three coded strings for item $i$, and let $h=16$ bytes be the framing cost per layer. The network-facing sizes are
\begin{equation}
B_i=\lvert b_i^z\rvert+\lvert b_i^A\rvert+h,
\qquad
E_i=\lvert b_i^N\rvert+h.
\label{eq:progressive-bytes}
\end{equation}
We retain each queued layer's next byte offset so interrupted service can resume in a later contact, and we admit refinement only after base completion.
\subsection{Codec-Native Semantic Value}

We estimate each item's clear-pixel count from the encoder-side parameters $s_i=h_s(\hat z_i)$ already computed while coding item $i$. With the codec fixed, we train a convolutional readout $g_\psi$ against $16\times16$ average-pooled binary cloud masks from AllClear training images. The predicted clear-pixel value is
\begin{equation}
v_i=HW\left(
1-\frac{1}{\lvert\Omega_s\rvert}
\sum_{u\in\Omega_s}\sigma\!\left(g_\psi(s_i)_u\right)
\right),
\label{eq:native-value}
\end{equation}
where $\Omega_s$ indexes the readout grid and $\sigma$ is the logistic function. We implement the readout with a $1\times1$ convolution from 384 to 128 channels, a $3\times3$ convolution from 128 to 64 channels, and a $1\times1$ scalar output, with rectified linear units after the first two layers. The readout has 123k parameters and requires 31.5\,M multiply-accumulate operations per tile.

We infer $v_i$ from $s_i$ once, without a separate onboard detector pass, fix it at arrival, and pass the contract $(a_i,B_i,E_i,v_i)$ to the network queue, where $a_i$ is item $i$'s arrival contact.

\subsection{Onboard Realization}
\label{sec:method:systems}

We optimize the onboard coded path. For each frame, we prepare one $256\times256$ tile, copy it to the accelerator, evaluate the analysis transform, hyperprior, checkerboard context, and entropy-parameter networks, copy symbols and probability indexes to the host, and entropy-code all streams. We exclude ground-side decoding from onboard service time. We replace per-symbol Gaussian-scale lookup with vectorized table indexing. Because the input size is fixed, we cache hyperlatent indexes and expanded medians, then replay the neural computation as a CUDA Graph after an explicit input copy~\cite{graphcompiler}.
We preserve the codec architecture, training loss, entropy coder, and bitstream syntax across the optimized and reference paths, yielding identical symbols, probability indices, and coded bytes in both paths. %Repeated 16-bit graph execution is deterministic.

%% file: sections/scheduling.tex
\section{Deadline-Pressure Contact Scheduling}\label{sec:scheduling}

With variable contact capacity, we must choose between new bases and refinements that complete older items. A fixed base allowance cannot respond to residual item work or aggregate deadline load. We rank arrived layers by semantic value, unfinished work, deadline slack, and the current queue's deadline pressure, and call this rule DPMW.

\subsection{Contact and Queue Model}

Item $i$ arrives at contact $a_i$ through the codec contract $(a_i,B_i,E_i,v_i)$. Its base and full deadlines are
\begin{equation}
d_i^B=a_i+1,
\qquad
d_i^F=a_i+3.
\label{eq:layer-deadlines}
\end{equation}
At contact $t$, we observe the current integer capacity $C_t$, and stored residual bytes cannot exceed $S_{\max}$. Let $x_{i\ell t}$ be bytes sent from layer $\ell\in\{B,E\}$ during contact $t$, where $B$ denotes the base layer and $E$ the refinement layer, let $r_{i\ell}(t)$ be its contact-start residual, and let $Q(t)$ be stored residual bytes. Feasible service satisfies
\begin{equation}
\begin{aligned}
\sum_{i,\ell}x_{i\ell t}&\leq C_t,
&
x_{i\ell t}&\in\{0,\ldots,r_{i\ell}(t)\},\\
Q(t)&\leq S_{\max}.
\end{aligned}
\label{eq:contact-feasibility}
\end{equation}
Let $\bar b_{\mathrm{CA},s}$ be the expected transmitted bytes per item for the full cloud-agnostic (CA) validation stream after weighting the data to 90\% mean cloud cover, including both 16-byte layer headers. For codec seed $s$ and $A=300$ arrivals per contact, one contact's capture volume is $V_s=\lfloor A\bar b_{\mathrm{CA},s}\rfloor$, and we set $S_{\max}=3V_s$. We retain the byte offset after partial service across contacts and admit refinement only after prior or same-contact base completion.

Let $T_i^F$ denote the full-completion contact. The full-delivery utility is
\begin{equation}
U_F=\sum_i v_i\,\mathds{1}\{T_i^F\leq d_i^F\}.
\label{eq:full-utility}
\end{equation}
Meeting $d_i^B$ is a separate usable-delivery objective, and we use one causal index policy for both. We compute operational weights from the native value $v_i$ and evaluate $\eta_B$ and $\eta_F$ with evaluator-only reference-clear pixels.

\subsection{Deadline Pressure}

At contact $t$, item $i$ has base residual $b_i(t)$, refinement residual $e_i(t)$, and full-delivery feasibility $f_i(t)\in\{0,1\}$, which becomes zero after either layer is evicted. The remaining feasible full-delivery work is
\begin{equation}
R_i^F(t)=f_i(t)\bigl(b_i(t)+e_i(t)\bigr).
\label{eq:full-chain-work}
\end{equation}
We compute $b_i(t)$, $e_i(t)$, $f_i(t)$, and $R_i^F(t)$ only from arrived items that remain stored at contact $t$.

At the start of each positive-capacity contact, we freeze the aggregate pressures
\begin{align}
P_B(t)
&=\sum_{\substack{i:b_i(t)>0\\t\leq d_i^B}}
\frac{b_i(t)}{d_i^B-t+1},
\label{eq:base-pressure}\\
P_F(t)
&=\sum_{\substack{i:R_i^F(t)>0\\t\leq d_i^F}}
\frac{R_i^F(t)}{d_i^F-t+1}.
\label{eq:full-pressure}
\end{align}
We interpret $P_B$ as the service rate required by incomplete bases and $P_F$ as that required by items whose full delivery remains feasible.

\subsection{Base and Refinement Weights}

For an incomplete base, the service weight is
\begin{equation}
\begin{split}
W_i^B(t)={}&
\mathds{1}\{t\leq d_i^B\}
\frac{v_i}{b_i(t)}
\frac{P_B(t)}{d_i^B-t+1}\\
&+
\mathds{1}\{R_i^F(t)>0,\ t\leq d_i^F\}
\frac{v_i}{R_i^F(t)}
\frac{P_F(t)}{d_i^F-t+1}.
\end{split}
\label{eq:base-weight}
\end{equation}
For an eligible refinement, the weight is
\begin{equation}
W_i^E(t)=
\mathds{1}\{R_i^F(t)>0,\ t\leq d_i^F\}
\frac{v_i}{R_i^F(t)}
\frac{P_F(t)}{d_i^F-t+1}.
\label{eq:refinement-weight}
\end{equation}
Through $v_i/b_i(t)$ and $v_i/R_i^F(t)$, we prioritize valuable items that require fewer residual bytes. The pressure-to-slack factors increase with queued work and approaching deadlines. 
%In $W_i^B$, we include one term for bases still eligible for usable delivery and one for base work required by feasible full deliveries, and in $W_i^E$ we include only the full-delivery term for eligible refinement.
\subsection{Causal Service Rule}

Arrivals enter before service. If they push stored residual bytes above $S_{\max}$, we evict refinement layers in increasing $v_i/e_i(t)$, followed by base layers in increasing $v_i/b_i(t)$. When we evict a base, we also remove its refinement.

For positive capacity, we freeze $P_B(t)$ and $P_F(t)$ and insert every positive-weight incomplete base and eligible refinement into a max heap. We serve the highest-weight layer until that layer or the contact capacity is exhausted. When a base completes, we immediately insert its refinement with the frozen pressures and the updated residual work $R_i^F(t)=e_i(t)$. After exhausting all positive-weight work, we serve expired work by decreasing native value per residual byte.

Eviction ties use earlier arrival and lower item identifier. Positive-weight ties use the earlier active deadline, then earlier arrival, base before refinement, and lower item identifier. Expired work ties omit the deadline key. For $L_t$ active layers, pressure construction takes $O(L_t)$ time, and heap service takes $O(L_t\log L_t)$ time and $O(L_t)$ memory. At each contact, we use current capacity, arrived item metadata, derived deadlines, residual bytes, and completion or eviction state. We use no future arrivals, interruption labels, evaluator-only clear pixels, fitted coefficients, capacity forecasts, or an offline solver.

\subsection{Clairvoyant Full-Delivery Bound}

We bound any scheduler on a fixed workload with a clairvoyant mixed-integer relaxation. Let $\mathcal K$ be the deadline-eligible cohort, $S_i=B_i+E_i$, $\mathcal{T}_i=\{a_i,\ldots,d_i^F\}$, and $n_i^C$ be evaluator-only reference-clear pixels. For a schedule $\pi$, define its evaluator full-delivery utility $U_F^C(\pi)=\sum_{i\in\mathcal K}n_i^C\mathds{1}\{T_i^F\leq d_i^F\}$. Binary $\zeta_i$ marks full completion, continuous $\phi_{it}$ denotes aggregate item bytes assigned to contact $t$, and $\xi_{it}=S_i \zeta_i-\sum_{t'=a_i}^{t-1}\phi_{it'}$ is selected residual work at contact start. We solve
\begin{equation}
\begin{aligned}
(U_F^C)^{\rm UB}=\max_{\zeta,\phi}\quad &\sum_{i\in\mathcal K}n_i^C\zeta_i\\
\text{s.t.}\quad
&\sum_{t\in \mathcal{T}_i}\phi_{it}=S_i \zeta_i &&\forall i,\\
&\sum_{i:t\in \mathcal{T}_i}\phi_{it}\leq C_t &&\forall t,\\
&\sum_{i:t\in \mathcal{T}_i}\xi_{it}\leq S_{\max} &&\forall t,\\
&\zeta_i\in\{0,1\},\quad \phi_{it}\geq0.
\end{aligned}
\label{eq:offline-bound}
\end{equation}
We give the model future arrivals and capacities, allow immediate eviction of unselected arrivals, omit unscored traffic, and drop the earlier base deadline. For any feasible system schedule $\pi$, retain only bytes of cohort items completed by $d_i^F$ and set their $\zeta_i=1$. Removing all other traffic cannot violate capacity or storage, so the retained service is feasible in~\eqref{eq:offline-bound} with the same evaluator utility. Thus $U_F^C(\pi)\leq U_F^{C,*}\leq (U_F^C)^{\rm UB}$, and $U_F^C(\pi)/(U_F^C)^{\rm UB}$ is a certified lower bound on its instance-specific ratio to the unknown clairvoyant optimum $U_F^{C,*}$. Aggregate service for a selected item has a base-first realization by labeling its first $B_i$ cumulative bytes as base. This certificate gives no worst-case approximation or stochastic stability guarantee.

%% file: sections/evaluation.tex
\section{Evaluation}\label{sec:eval}

\subsection{Experimental Design}

We fit models on AllClear training data. We use validation data to select operating points and network resources and to report the progressive-byte and 501-frame deletion comparisons. We use the region-disjoint test split for rate-distortion, per-image, capacity-response, and scheduler results. Each split contains $256\times256$ Sentinel-2 crops and machine-generated cloud masks~\cite{allclear}. We test unchanged same-sensor transfer on expert-labeled CloudSEN12+~\cite{cloudsen12plus}. We train four codec seeds at five distortion weights and use three shared rate points for transfer and scoring sensitivity. We count emitted range asymmetric numeral system (rANS) bytes from 32-bit floating-point encoding of every image and hyperlatent stream and compute Bj{\o}ntegaard delta rate (BD-rate) at matched clear-region PSNR with 95\% Student-$t$ intervals over codec seeds. We retrain ELIC~\cite{elic} and Minnen 2018~\cite{mbt2018mean} from scratch on the same training data with four seeds and five distortion weights per codec spanning a comparable rate range with one shared early-stopping schedule. 
%We encode JPEG 2000 with OpenJPEG at nine fixed ratios and exclude the one point below 38\,dB clear-region PSNR from BD-rate.

We measure encoder and standalone-detector execution on a Jetson Orin Nano Super and Jetson AGX Orin. For the primary Nano configuration, we use 16-bit floating point, batch size one, 15\,W mode, and maximum clocks without TensorRT. We compute linearly interpolated p99 from 100 synchronized calls after 30 warm-ups and subtract separately measured idle module-rail power. Before test evaluation, we use validation data to select one clear-weighted (CW) point and one quality-matched CA point. We encode each capture into base and refinement layers with 16-byte headers while retaining cross-contact state.

For the orbit-profile experiment, we draw 300 arrivals per contact from 4,052 test frames resampled to 90\% mean cloud cover. We use one-contact base and three-contact full deadlines, packet resumption, five backlog orders per codec seed, and seven capacity fractions. We propagate a 24-hour Sentinel-2B orbit over a Svalbard ground station with a $5^\circ$ elevation mask and set the 15 relative contact capacities to the contacts' usable byte volumes. Each backlog order is one seeded arrival draw. At fraction $f$, we distribute $\lfloor15fV_s\rfloor$ bytes across contacts according to the orbit profile. We hold arrivals, integer capacities, storage limits, and backlog order fixed between paired CA and CW cells. We repeat the sweep with native, shared-detector, and uniform values. With fixed two-stage service, we maximize complete due-base native value before admitting refinement while retaining at least 99\% of that optimum. We average capacity curves within codec seed and apply one-sided Student-$t$ inference to the normalized capacity-range integral.
For the held-out scheduler experiment, we reuse the four selected CW streams over 30 contacts. We evaluate five backlog orders, two capacity fractions, and three interruption events in 240 paired interrupted and matched-control cells. For each event, we set one to three consecutive contact capacities to zero and either redistribute the removed bytes over the next six contacts or remove them permanently. We give the matched controls the same total bytes on the uninterrupted profile. Only byte-service order changes across the three schedulers. 
%We define the combined cohort from three contacts before interruption onset through six contacts after the last interrupted contact.
We score the combined cohort, the items arriving from three contacts before the first interrupted contact through six after the last one. Earlier arrivals reach their full deadlines before the interruption, and the six-contact tail spans the redistribution window.
For deadline-full delivery $\eta_F$, we credit reference-clear pixels in items whose base and refinement meet the three-contact deadline. For deadline-usable delivery $\eta_B$, we use the same denominator and credit a completed base at one contact. We average balanced cells within codec seed and compute one-sided 95\% lower bounds and interruption-specific difference-in-differences (DiD) conditional on the four seeds. In post hoc replays, we apply four online orders with identical state-transition rules: earliest deadline first (EDF) with native value per residual byte as tiebreak, native value, native value per residual full-delivery work, and that ratio divided by deadline slack. We solve~\eqref{eq:offline-bound} with HiGHS at relative gap $10^{-6}$.

\subsection{Source Representation and Allocation}

\subsubsection{Capture-Time Deletion}

\begin{figure}[b]
\centering
\includegraphics[width=\columnwidth]{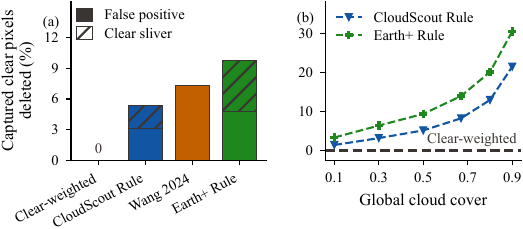}
\caption{Clear-pixel deletion from frame- and pixel-level cloud removal.}
\label{fig:deletion}
\end{figure}

We apply one fixed standalone detector to the CloudScout 70\% and Earth+ 50\% frame rules~\cite{cloudscout,earthplus} on a 501-frame trace and sweep cloud cover through weighted resampling with five 4,000-frame backlogs. We separate deletion caused by frame-level false positives from deletion of clear slivers, the clear regions inside correctly discarded cloudy frames, in Fig.~\ref{fig:deletion}.

We measure 5.36\% clear-pixel deletion with the CloudScout Rule on the fixed trace. Frame-level false positives account for 3.08 points and clear slivers for 2.28 points. With the Earth+ Rule, we measure 9.76\% deletion. Our pixel-removal adaptation of Wang et al. deletes 7.36\%~\cite{wang2024}. At 90\% mean cover, deletion reaches $21.4\pm1.7$\% for the CloudScout Rule and $30.4\pm1.8$\% for the Earth+ Rule.

We transfer the fixed detector and both codecs to CloudSEN12+ without retraining (Table~\ref{tab:cs12}).
\input{tables/tab_cs12}

Both frame rules delete more than one-fifth of the captured clear pixels on CloudSEN12+, primarily due to false positives.

\input{sections/evaluation_compression}

\subsection{Deadline-Constrained Delivery}

\subsubsection{Progressive Capacity Response}

At the selected operating points, CW transmits 21\% fewer bytes than CA, including layer headers, on the full 501-frame validation split without cloud-cover resampling.
\begin{figure}[t]
\centering
\includegraphics[width=\columnwidth]{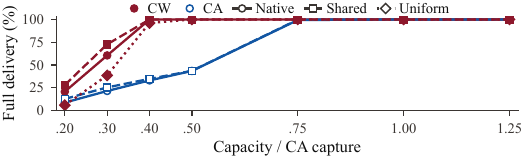}
\caption{Deadline-full delivery across contact capacities for fixed two-stage service.}
\label{fig:progressive-capacity}
\end{figure}

With the same fixed two-stage service and codec-native values, we measure a 20.25-point higher normalized capacity-range integral for CW than for quality-matched CA, with a 17.41-point lower bound (Fig.~\ref{fig:progressive-capacity}). With the shared-detector score, we measure a 20.93-point increase. When we replace uniform with native values, the CW integral increases by 3.16 points. We reach 95\% mean deadline-full delivery at about 47\% lower capacity with CW than with CA for both native and shared scores. The validation-selected CW delivery point lies above the plotted rate range of Fig.~\ref{fig:rd}. Across the four CW test streams, scored without cloud-cover resampling, we measure a base-layer fraction of 66.0\% of mean full wire bytes and base clear-region PSNR of 44.63\,dB, 1.88\,dB below full reconstruction.

\subsubsection{Deadline-Pressure Scheduling}

For deadline interleaving, we serve due bases, due refinements, future bases, and future refinements in fixed phases and order each phase by deadline and native value per residual byte.
\begin{figure}[t]
\centering
\includegraphics[width=\columnwidth]{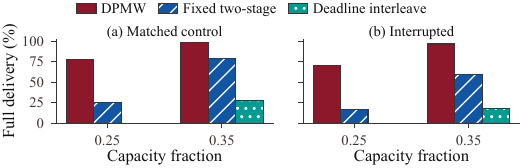}
\caption{Combined-cohort deadline-full delivery for DPMW and both online baselines.}
\label{fig:dpmw-capacity}
\end{figure}

DPMW raises mean deadline-full delivery from 38.1\% to 83.5\% versus fixed two-stage service and by 74.77 points versus deadline interleaving (Fig.~\ref{fig:dpmw-capacity}, Table~\ref{tab:dpmw-heldout}).
\input{tables/tab_dpmw_heldout}

The DPMW--fixed gain is 45.41 points, with a 35.95-point lower bound. We compute a post hoc paired cluster bootstrap over codec seeds and backlog orders, stratified by event and capacity cell, and obtain a two-sided 95\% interval of 37.31--50.21 points. All three schedulers transmit 99.72\% of available capacity on average across interrupted profiles, so service volume does not explain the delivery difference. Deadline interleaving keeps a base in the due phase after its deadline passes and serves due bases oldest first, so expired bases precede just-due ones and due refinements wait behind the whole due-base backlog.

\begin{figure}[t]
\centering
\includegraphics[width=\columnwidth]{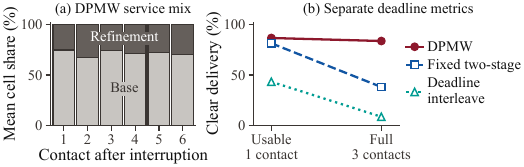}
\caption{DPMW service mix and combined-cohort delivery at two deadlines.}
\label{fig:dpmw-mechanism}
\end{figure}

With DPMW across the six post-interruption contacts, we allocate 25--33\% of transmitted bytes to refinement while continuing base service (Fig.~\ref{fig:dpmw-mechanism}). Mean full and usable delivery remain closer with DPMW than with either online baseline.

\input{tables/tab_dpmw_references}

DPMW raises mean full delivery by 11.30 points and mean usable delivery by 6.28 points relative to EDF on interrupted profiles (Table~\ref{tab:dpmw-references}). The two density policies exceed DPMW's full delivery by 2.13--2.21 points but fall 2.44--3.21 points below its usable delivery. We found no improvement from the aggregate-pressure multipliers in post hoc ablations. DPMW reaches 83.60\% of the certified bound and closes 73.49\% of the fixed-to-bound full-delivery gap. The certified bound completes about one-third of cohort items at the lower capacity fraction, and the uncompleted items hold under 1\% of cohort clear pixels. HiGHS reports optimal status for all 240 solves.
During interruptions, deadline-usable delivery is 5.22 points higher with DPMW than with fixed two-stage service, with a 2.88-point lower bound. The full-delivery improvement is 9.66 points larger for interrupted profiles than for matched controls, with an 8.43-point lower bound. In matched controls, full delivery is higher with DPMW, while one-contact usable delivery is 1.99 points lower.

\subsection{Onboard Cost}\label{sec:eval:systems}

\input{tables/tab_deadline_v3}
Optimizing the coding path reduces fused 99th-percentile service time from 47.03 to 19.89\,ms and idle-subtracted energy from 148.25 to 103.17\,mJ per frame on the primary Nano configuration (Table~\ref{tab:deadline}). Optimized component medians remain below 33\,ms on both devices.

On 4,052 held-out tiles per codec seed, the native readout has a mean Spearman correlation of 0.852 and a mean absolute error of 6,820 pixels, or 10.41\% of a tile. The separate detector reaches 0.887 and 6,131 pixels, or 9.35\%. 
%We have not measured native-readout latency or energy, and retain the detector only as a cost baseline.
%
In separate active windows, detector-to-encoder energy ratios for the full encoder are 2.21--2.62 across device and power configurations.
\begin{figure}[t]
\centering
\includegraphics[width=\columnwidth]{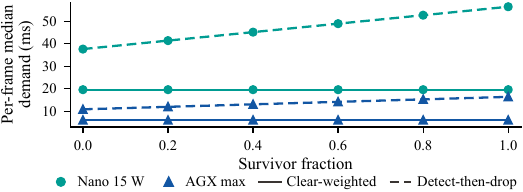}
\caption{Per-frame service demand across survivor fractions on two devices.}
\label{fig:survivor}
\end{figure}

\begin{figure}[t]
\centering
\includegraphics[width=\columnwidth]{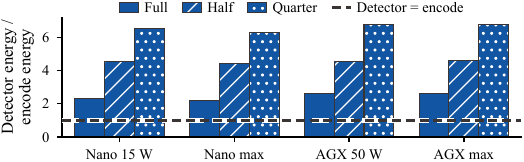}
\caption{Detector energy divided by optimized encoder energy at maximum clocks.}
\label{fig:energy}
\end{figure}

On the Nano, detection takes 36.9\,ms versus 19.6\,ms for full-frame encoding, and detect-then-drop remains slower than encoding every frame across both device curves (Fig.~\ref{fig:survivor}).
For the energy-only size comparison, the full, half, and quarter encoders contain 15.9, 7.7, and 4.2 million onboard parameters. Across twelve maximum-clock device, power, and encoder-size configurations, detector-to-encoder energy ratios span 2.21--6.78 (Fig.~\ref{fig:energy}).

%% file: tables/tab_cs12.tex
\begin{table}[b]
  \centering
  \caption{Capture-time deletion and frozen transfer to CloudSEN12+.}
  \label{tab:cs12}
  \scriptsize
  \setlength{\tabcolsep}{4pt}
  \setlength{\aboverulesep}{0.15ex}
  \setlength{\belowrulesep}{0.35ex}
  \begin{tabular*}{\columnwidth}{@{\extracolsep{\fill}}lcc@{}}
    \toprule
    Metric & AllClear & CloudSEN12+ \\
    \midrule
    \multicolumn{3}{@{}l}{\emph{Deletion, total \% (false-positive points)}} \\
    CloudScout Rule & 5.36 (3.08) & 21.19 (16.95) \\
    Earth+ Rule & 9.76 (4.73) & 28.74 (17.80) \\
    Clear-weighted vs.\ CA BD-rate (\%) & $-37.6 \pm 4.6$ & $-26.4 \pm 5.0$ \\
    \bottomrule
  \end{tabular*}
\end{table}

%% file: sections/evaluation_compression.tex
\subsubsection{Matched Clear-Region Quality}

\begin{figure}[t]
\centering
\includegraphics[width=\columnwidth]{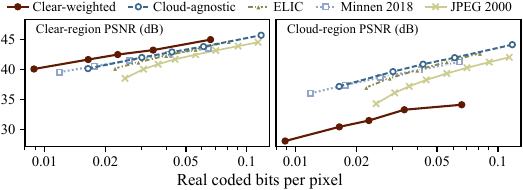}
\caption{Clear- and cloud-region rate-distortion on AllClear.}
\label{fig:rd}
\end{figure}

Clear-weighted coding reduces real bytes by 39.0\% at matched clear-region quality against the same architecture trained without clear weighting (Fig.~\ref{fig:rd}). Relative to Efficient Learned Image Compression (ELIC), we measure 47.8\% fewer bytes, with fewer bytes for every trained seed than for either reference. Cloud-agnostic training with our internal architecture reduces bytes by 13.7\% relative to ELIC. Minnen 2018 is the stronger learned reference at the evaluated rate points (Table~\ref{tab:bdrate}).
\input{tables/tab_c1_bdrate}

At 0.0164\,bpp, clear-region PSNR is 1.5\,dB higher and cloud-region PSNR is 6.7\,dB lower with clear weighting. Against the same cloud-agnostic codec over the three shared rate points, we measure a 37.6\% byte reduction on AllClear and 26.4\% after unchanged transfer of both codecs to CloudSEN12+ (Table~\ref{tab:cs12}).

\subsubsection{Conditional Codelength}

\begin{figure}[t]
\centering
\includegraphics[width=\columnwidth]{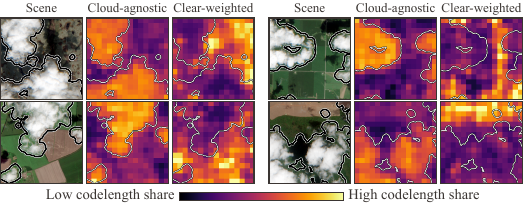}
\caption{Normalized image-latent codelength for four partly cloudy scenes.}
\label{fig:maps}
\end{figure}

We compare normalized image-latent codelength for four partly cloudy scenes in Fig.~\ref{fig:maps}.

At distortion weight 300, we allocate $5.09\pm0.51$ times as many image-latent bits per clear pixel as per cloudy pixel with clear-weighted coding. For cloud-agnostic coding, the ratio is $0.52\pm0.02$. We compute conditional image-latent codelength from the exact rANS cumulative-distribution tables, exclude hyperlatents and stream flush, and normalize each map to within-image codelength shares.

\subsubsection{Per-Image Effects}

\begin{figure}[t]
\centering
\includegraphics[width=\columnwidth]{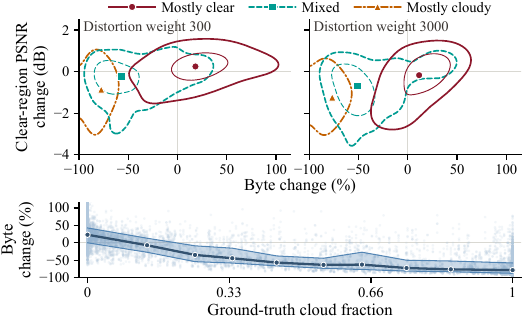}
\caption{Per-image real-byte and clear-region PSNR changes across cloud cover at distortion weights 300 and 3000.}
\label{fig:perimage}
\end{figure}

We report byte and clear-region PSNR changes across reference cloud fraction in Fig.~\ref{fig:perimage}.

We define mostly clear and mostly cloudy groups using the lower and upper thirds of reference cloud fraction. At distortion weight 300, mostly clear scenes use 18.2\% more bytes at the median and gain 0.25\,dB median clear-region PSNR. Mostly cloudy scenes use 77.7\% fewer bytes, with a 0.86\,dB median reduction among scenes containing scored clear pixels. At weight 3000, mostly clear scenes use 11.5\% more bytes while median clear-region PSNR falls by 0.17\,dB.

\subsubsection{Scoring Sensitivity}

We compare BD-rate with ELIC using four clear-region definitions in Table~\ref{tab:region}.
\input{tables/tab_c2_region}

For every definition, we measure at least 43\% fewer bytes with clear-weighted coding than with ELIC at the three shared rate points, while the cloud-agnostic reduction remains near 16\%. When we exclude clear pixels within one and two pixels of a cloud boundary, the clear-weighted reduction increases. With the population scored by OmniCloudMask, we still measure fewer bytes with clear-weighted coding than with cloud-agnostic coding~\cite{omnicloudmask}.

%% file: tables/tab_c1_bdrate.tex
\begin{table}[b]
  \centering
    \caption{Real-byte BD-rate at matched clear-region PSNR over five rate points.}
  \label{tab:bdrate}
  \scriptsize
  \setlength{\tabcolsep}{3pt}
  \setlength{\aboverulesep}{0.15ex}
  \setlength{\belowrulesep}{0.35ex}
  \begin{tabular*}{\columnwidth}{@{\extracolsep{\fill}}llr@{\,$\pm$\,}r@{}}
    \toprule
    Method & Reference & \multicolumn{2}{r@{}}{\makebox[0pt][r]{BD-rate (\%)}} \\
    \midrule
    \multirow{4}{*}{Clear-weighted} & Cloud-agnostic & $-39.0$ & $7.4$ \\
                                       & ELIC                    & $-47.8$ & $7.4$ \\
                                       & Minnen 2018             & $-42.0$ & $6.4$ \\
                                       & JPEG 2000               & $-62.2$ & $4.7$ \\
    \midrule
    Cloud-agnostic & ELIC                    & $-13.7$ & $4.1$ \\
    \bottomrule
  \end{tabular*}
\end{table}

%% file: tables/tab_c2_region.tex
\begin{table}[t]
  \centering
  \caption{Real-byte BD-rate (\%) against ELIC at the three shared rate points.}
  \label{tab:region}
  \scriptsize
  \setlength{\tabcolsep}{3.2pt}
  \setlength{\aboverulesep}{0.15ex}
  \setlength{\belowrulesep}{0.35ex}
  \begin{tabular*}{\columnwidth}{@{\extracolsep{\fill}}lrrrr@{}}
    \toprule
    Model & AllClear & $+1$ px & $+2$ px & Omni \\
    \midrule
    Clear-weighted & $-50.2\pm6.4$ & $-51.5\pm6.7$ &
    $-52.7\pm6.6$ & $-43.7\pm9.6$ \\
    Cloud-agnostic & $-16.0\pm4.5$ & $-15.9\pm4.3$ &
    $-16.0\pm4.3$ & $-16.2\pm5.0$ \\
    \bottomrule
  \end{tabular*}
\end{table}

%% file: tables/tab_dpmw_heldout.tex
% provenance: CONTRIBUTION-BUNDLE.json and heldout-v1/selection.json
\begin{table}[b]
  \centering
  \caption{Combined-cohort DPMW contrasts.}
  \label{tab:dpmw-heldout}
  \scriptsize
  \setlength{\tabcolsep}{3.2pt}
  \setlength{\aboverulesep}{0.15ex}
  \setlength{\belowrulesep}{0.35ex}
  \resizebox{\columnwidth}{!}{%
  \begin{tabular}{@{}lrr@{}}
    \toprule
    Contrast & Mean (points) & 95\% lower bound (points) \\
    \midrule
    Interrupted full, vs. fixed & $+45.41$ & $+35.95$ \\
    Interrupted full, vs. interleave & $+74.77$ & $+65.72$ \\
    Interruption-specific full DiD & $+9.66$ & $+8.43$ \\
    Interrupted usable, vs. fixed & $+5.22$ & $+2.88$ \\
    \bottomrule
  \end{tabular}}
\end{table}

%% file: tables/tab_dpmw_references.tex
% provenance: INFOCOM-DPMW-ABLATION and INFOCOM-DPMW-ORACLE
\begin{table}[t]
  \centering
  \caption{Post hoc references on interrupted profiles.}
  \label{tab:dpmw-references}
  \scriptsize
  \setlength{\tabcolsep}{3.2pt}
  \setlength{\aboverulesep}{0.15ex}
  \setlength{\belowrulesep}{0.35ex}
  \begin{tabular}{@{}lrr@{}}
    \toprule
    Policy or bound & $\eta_B$ (\%) & $\eta_F$ (\%) \\
    \midrule
    Clairvoyant upper bound & -- & $99.89$ \\
    Density / deadline slack & $84.04$ & $85.72$ \\
    Value / residual full work & $83.27$ & $85.64$ \\
    Native value & $80.74$ & $83.73$ \\
    DPMW & $86.47$ & $83.51$ \\
    EDF & $80.20$ & $72.21$ \\
    \bottomrule
  \end{tabular}
\end{table}

%% file: tables/tab_deadline_v3.tex
\begin{table}[t]
\centering
\caption{Onboard encoder service time and energy.}
\label{tab:deadline}
\scriptsize
\setlength{\tabcolsep}{2.6pt}
\setlength{\aboverulesep}{0.15ex}
\setlength{\belowrulesep}{0.35ex}
\begin{tabular*}{\columnwidth}{@{\extracolsep{\fill}}lllrrr@{}}
\toprule
\multicolumn{6}{@{}l}{\emph{Fused full encoder, Orin Nano 15 W, max clocks}} \\
Execution
& \multicolumn{2}{c}{p99 (ms)}
& 33-ms margin
& \multicolumn{2}{r@{}}{Energy (mJ)} \\
\midrule
Reference & \multicolumn{2}{c}{47.03} & $-14.03$
& \multicolumn{2}{r@{}}{148.25} \\
Optimized & \multicolumn{2}{c}{19.89} & $+13.11$
& \multicolumn{2}{r@{}}{103.17} \\
\midrule
\multicolumn{6}{@{}l}{\emph{Component medians, reference$\rightarrow$optimized}} \\
Device & Mode & Encoder
& Service time (ms)
& Margin (ms)
& Energy (mJ) \\
\midrule
Nano & 15 W & full & $46.0\rightarrow19.6$ & 13.4 & $148\rightarrow103$ \\
\midrule
AGX & max & full & $23.6\rightarrow6.0$ & 27.0 & $36\rightarrow18$ \\
\bottomrule
\end{tabular*}
\end{table}

%% file: sections/related_work.tex
\section{Related Work}\label{sec:related}

Onboard Earth-observation systems reduce downlink demand through frame or tile rejection, cloud filling, adaptive coding and reconstruction, and re-imaging schedules~\cite{cloudscout,kodan,earthplus,wang2024,deepspace,scocloud}. Zhang et al. select and deduplicate tiles for object counting in TargetFuse, while Liu et al. remove spatial and temporal redundancy from space-situational-awareness imagery through distributed on-orbit sparse coding~\cite{DBLP:conf/infocom/ZhangYXZZMXDW24,DBLP:conf/infocom/LiuJYCZKLLC25}. These systems either make content decisions outside the codec or exploit generic image redundancy. We use dense cloud probabilities only during training to allocate rate within every RGB capture. During operation, we encode without a separate onboard detector pass and transmit no cloud map. We expose resumable base and refinement bytes to the scheduler.

Prior studies optimize satellite codecs for global, metadata-conditioned, temporal, or task-specific reconstruction~\cite{phisat2,cosmic,terracodec,rs_llic,molliere2025}. Other authors allocate rate through importance maps, training masks, segmentation priors, or global distortion~\cite{cwic,colic,segpic,elic,mbt2018mean}. SPIFF, AdaStreamer, AdaSem, and OMNIS adapt imagery or features to semantic relevance, bandwidth, channel feedback, or radio resources~\cite{DBLP:conf/infocom/PalenaAGC26,DBLP:conf/infocom/Zhu0CSQL24,DBLP:conf/infocom/LiaoT24,DBLP:conf/infocom/QinHNCL26}. These systems infer relevance during operation or bind communication to a fixed task. We retain general image reconstruction and export exact layer sizes with a codec-native value.

Liu et al. assign tasks to local, sunlit, or ground resources in PHOENIX to meet completion deadlines while minimizing maximum battery depth of discharge~\cite{DBLP:conf/infocom/LiuLWLZLLL24}. Park et al. coordinate satellite handoff and video rate from throughput predictions to maximize streaming quality~\cite{DBLP:conf/infocom/ParkHLXQGM26}. Neither group allocates contact bytes among incomplete base and refinement layers from multiple captures.
We rank residual base and refinement bytes with DPMW while limiting storage, enforcing distinct deadlines, resuming packets, and using no future-capacity forecast.

Our earlier work, FOOL~\cite{fool}, pairs feature compression via shallow variational bottleneck injection~\cite{frankensplit} with accounting for target-device throughput, transfer cost, energy, and mission capacity. There, we deliberately set clouds aside as a separate challenge and preserved downstream task performance without prior knowledge of the tasks. In contrast, this work addresses the more general objective of preserving clear-region reconstruction quality, saving the bits that cloud regions would otherwise consume as high-entropy content. We additionally consider capture-time deletion, progressive coded bytes, and deadline delivery.
%Furutanpey et al. pair task-agnostic feature compression from shallow variational bottleneck injection~\cite{frankensplit} with target-device throughput, transfer cost, energy, and mission accounting in FOOL~\cite{fool}.
%We follow the paired compression-and-systems evaluation and additionally measure encoder service time, capture-time deletion, progressive coded bytes, and deadline delivery.

%% file: sections/conclusion.tex
\section{Conclusion}\label{sec:conclusion}

Clear-weighted training reallocates source bits toward clear content and emits resumable base and refinement layers with codec-native values. DPMW ranks arrived layers by value, residual work, and two deadlines while limiting storage, resuming packets, and using no future-capacity forecast. The selected progressive stream uses 21\% fewer bytes than the cloud-agnostic stream from the same codec architecture on the validation split. In held-out interrupted contacts, DPMW raises deadline-full delivery from 38.1\% to 83.5\%, with a 35.95-point lower bound. It reaches 83.6\% of the certified clairvoyant upper bound and exceeds the four replayed reference orders in deadline-usable delivery. The optimized encoder stays below the latency target on the primary Nano configuration.
Future work will bridge this system with FOOL's task-agnostic feature compression and extend the clear-weighted objective and the scheduler to preserve performance on downstream tasks without prior knowledge of their labels.
%Mission-specific base-versus-full utility design and complete arrival-to-bitstream measurements, including the native readout, remain open.